\documentclass[aps,prl,superscriptaddress,reprint]{revtex4-2}
\usepackage{centernot}
\usepackage{graphicx}
\usepackage{amsmath}
\usepackage{times}
\usepackage{amssymb}
\usepackage{mathrsfs}
\usepackage{chemarr}
\usepackage{xcolor}
\usepackage{url}
\usepackage{version}
\newcommand{\br}{\textbf{r}}

\graphicspath{{./Image/}}

\usepackage[pdftex,colorlinks=true,pdfstartview=FitV,linkcolor= linkcolor,citecolor= linkcolor,urlcolor= linkcolor,hyperindex=true,hyperfigures=false]{hyperref}
\definecolor{linkcolor}{rgb}{0,0,0.6}

\usepackage{lipsum}
\newcommand\bfr{{\bf r}}
\newcommand\bfj{{\bf j}}
\newcommand\bfu{{\bf u}}
\newcommand\bfF{{\bf F}}
\newcommand\bfeta{{\boldsymbol{\eta}}}

\newcommand\bfq{{\bf q}}
\newcommand\bfm{{\bf m}}
\newcommand\bfM{{\bf M}}
\usepackage{tikz}

\newcommand\mv[1]{ \langle #1 \rangle}

\newcommand\bfI{{\bf I}}
\newcommand\bfJ{{\bf J}}
\newcommand\bfQ{{\bf Q}}
\newcommand\bfchi{{\boldsymbol \chi}}
\newcommand\mass{\mathfrak{m}}
\newcommand\bsigma{\boldsymbol{\sigma}}

\begin{document}


\title{Nematic Torques in Scalar Active Matter: When Fluctuations Favor Polar Order and Persistence}

\author{Gianmarco Spera}
\affiliation{Université Paris Cité, Laboratoire Mati\`ere et Syst\`emes Complexes (MSC), UMR 7057 CNRS,  F-75205 Paris,  France}

\author{Charlie Duclut}
\affiliation{Université Paris Cité, Laboratoire Mati\`ere et Syst\`emes Complexes (MSC), UMR 7057 CNRS,  F-75205 Paris,  France}
\affiliation{Laboratoire Physique des Cellules et Cancer (PCC), CNRS UMR 168, Institut Curie, Université PSL, Sorbonne Université, 75005 Paris, France}

\author{Marc Durand}
\affiliation{Université Paris Cité, Laboratoire Mati\`ere et Syst\`emes Complexes (MSC), UMR 7057 CNRS,  F-75205 Paris,  France}

\author{Julien Tailleur}
\affiliation{Department of Physics, Massachusetts Institute of Technology, Cambridge, Massachusetts 02139, USA}
\affiliation{Université Paris Cité, Laboratoire Mati\`ere et Syst\`emes Complexes (MSC), UMR 7057 CNRS,  F-75205 Paris,  France}

\date{\today}

  \begin{abstract}
 We study the impact of nematic alignment on scalar active matter in
 the disordered phase. We show that nematic torques control the
 emergent physics of particles interacting via pairwise forces and can
 either induce or prevent phase separation. The underlying
 mechanism is a fluctuation-induced renormalization of the mass of the
 polar field that generically arises from nematic torques. The
 correlations between the fluctuations of the polar and nematic fields
 indeed conspire to increase the particle persistence length, contrary
 to what phenomenological computations predict. This effect is generic
 and our theory also quantitatively accounts for how nematic torques
 enhance particle accumulation along confining boundaries and opposes
 demixing in mixtures of active and passive particles.
\end{abstract}

\maketitle

Active matter describes systems comprising elementary units able to
exert non-conservative forces on their
environment~\cite{marchetti2013hydrodynamics,bechinger2016active,chate2020dry,o2022time}. Such
systems are ubiquitous in
nature~\cite{Ballerini2008,schaller2010polar,battle2016broken,poujade2007collective}
and can also be engineered in the
lab~\cite{howse2007self,bricard2013emergence,nishiguchi2015mesoscopic,yan2016reconfiguring},
 paving the way towards the engineering of soft active
materials. A key requirement to do so is the ability to predict how
microscopic characteristics of active systems impact their emerging
behaviors. However, we still lack a statistical mechanics treatment generalizing what is classically done for equilibrium
systems.

Following pioneering works~\cite{vicsek1995novel,toner1995long},
progress has been made over the past fifteen years to account for the
large-scale properties of `simple' active systems, in which
self-propulsion interplays with a single other ingredient, be it
aligning torques~\cite{ngo2014large,solon2015phase,chate2020dry}, external
potentials~\cite{galajda2007wall,tailleur2009sedimentation,di2010bacterial,sokolov2010swimming,takatori2016acoustic,baek2018generic},
pairwise
forces~\cite{fily2012athermal,redner2013structure,stenhammar2014phase,wysocki2014cooperative,cates2015motility,solon2018generalized,digregorio2018full,caporusso2020motility,omar2021phase,speck2021coexistence},
or mediated
interactions~\cite{saintillan2008instabilities,tailleur2008statistical,saha2014clusters,zakine2018field}. However,
realistic systems typically involve many of these aspects
simultaneously and the resulting physics is both much richer and much
harder to account
for~\cite{deseigne2010collective,schaller2011frozen,peruani2012collective,sumino2012large,geyer2019freezing}. In
particular, the interplay between pairwise forces and aligning torques
has attracted a lot of interest
recently~\cite{peruani2011traffic,farrell2012pattern,weber2013long,sese2018velocity,shi2018self,van2019interparticle,van2019interrupted,grossmann2020particle,sese2021phase,chate2020dry}.

Aligning interactions obviously play a crucial role in the ordered
phases they induce, where they lead to a wealth of dynamical patterns
that have been extensively studied~\cite{chate2020dry}. Their impact
in the disordered phase, on the contrary, remains largely
unexplored. For concreteness, we work in $d=2$ dimensions and consider
$N$ active particles of positions $\bfr_i$ and orientations
$\bfu_i=(\cos\theta_i,\sin\theta_i)$ evolving as
\begin{subequations}\label{eq:dynamics}
\begin{align}
  \dot \bfr_i &= v_0 \bfu_i +\mu \sum_{\langle j ,i\rangle} \bfF_{ji}+\sqrt{2
    D_t}\bfeta_i \label{eq:dynamicspos}\\ \dot \theta_i &=
  \frac{\gamma}{n_i} \sum_{\langle j, i\rangle} \sin[p(\theta_j-\theta_i)]+\sqrt{2
    D_r}\zeta_i\label{eq:dynamicsangle}\;,
\end{align}
\end{subequations}
where $v_0$ is the particle self-propulsion velocity, $\mu$ is the
particle mobility, $D_t$ is the translational diffusivity, $\bfeta_i$
and $\zeta_i$ are centered unit-variance Gaussian white noises, and
$\bfF_{ji}$ is the force exerted by particle $j$ onto
particle~$i$. Particles interact when their distance is smaller than
$r_0$ and $n_i=\sum_{\langle j ,i\rangle} 1$ is the number of
particles interacting with particle $i$. One usually refers to the
aligning interactions as polar when $p=1$ and nematic when $p=2$,
consistent with the symmetry of the ordered phase they favor.
Finally, $\gamma$ is the alignment strength and $D_r$ is the rotational diffusivity, which controls the
microscopic persistence time $\tau=D_r^{-1}$.

The role of aligning interactions in the disordered phase can be
analyzed from symmetry considerations. Particle conservation makes the
density field a hydrodynamic mode, whose evolution is governed by a
density current: $\partial_t \rho(\bfr,t)=-\nabla \cdot
\bfj(\bfr,t)$. The latter includes an advective contribution $v_0
\bfm$ due to self-propulsion, where $\bfm(\bfr,t)$ is the orientation
field of the particles. Since polar torques ($p=1$) align the particle
orientations, they control the particle
persistence~\cite{sese2018velocity} and lead to an evolution for
$\bfm$ reminiscent of a Landau theory:
  $\partial_t \bfm = -(T-T_c) \bfm + [\dots]$,
where $[\dots]$ refers to transport terms and higher-order
contributions. The collective persistence time of the advective
current due to self-propulsion thus scales as $\tau^c=(T-T_c)^{-1}$ in
the disordered (or `high-temperature') phase: polar torques enhance
persistence by decreasing the `mass' (that is to say, the large-scale
limit of the inverse relaxation time of fluctuations) of the polar
field. Immediately, this implies that aligning interactions can
promote collective behaviors such as motility-induced phase separation
(MIPS)~\cite{tailleur2008statistical,fily2012athermal,redner2013structure,stenhammar2014phase,wysocki2014cooperative,cates2015motility,solon2018generalized,digregorio2018full,caporusso2020motility,omar2021phase,speck2021coexistence}
by lowering the critical `bare' persistence length $v_0/D_r$ above
which MIPS can be observed~\cite{sese2018velocity,sese2021phase}.

Nematic torques ($p=2$) are ubiquitous among active particles since
polar shapes generically lead to nematic alignment upon
collisions. The symmetry argument presented above for the polar case
does not predict anything interesting for the nematic case: to leading
order, the Landau theory reads $\partial_t \bfm = -D_r \bfm + \kappa
\bfm \cdot \bfq + [\dots]$, where $\bfq$ is the nematic order
parameter and $\kappa$ is proportional to the amplitude of the nematic
torques. In the high-temperature phase, $\bfq=0$ and mean-field theory
predicts that nematic torques do not impact the particle
persistence~\cite{sese2021phase}, hence leaving their emerging
behaviors unaltered. While the ordered phases induced by nematic
torques have attracted a lot of
attention~\cite{simha2002hydrodynamic,toner2005hydrodynamics,ginelli2010large,nishiguchi2017long,shankar2018low,chate2020dry,mahault2021long},
the impact of nematic alignment on disordered scalar active matter has
thus been little studied.

\begin{figure}
  \begin{center}
    \includegraphics{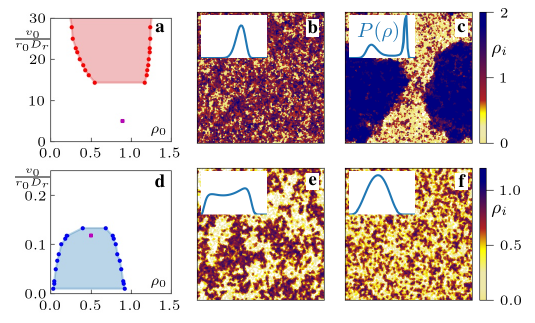}
  \end{center}
  \caption{Impact of nematic torques on the phase separation of active
    particles evolving under Eq.~\eqref{eq:dynamics}. Particles
    interact via either purely repulsive Weeks-Chandler-Andersen
    potential {(top)} or Lennard-Jones potential {(bottom)}. {\bf
      (a,d)} Phase diagrams as the average density $\rho_0$ and the
    rotational diffusivity $D_r$ are varied, in absence of nematic
    torques. Shaded regions correspond to phase-separated systems and
    are delimited by coexistence lines (symbols).  {\bf (b,e)}
    Snapshots corresponding to the magenta squares in (a,d). {\bf
      (c,f)} Snapshots obtained from (b,e) upon the addition of
    nematic torques: $\gamma=6.5 D_r$ in (c) and $\gamma=5.9 D_r$ in
    (f). The color of particle $i$ encodes $\rho_i=n_i/(\pi r_0^2)$.
    Global nematic order is observed for $\gamma\gtrsim 7
    D_r$. Insets in (b,c,e,f) show histograms of the local
    density. Bimodality signals phase separation. Nematic torques
    induce MIPS in (c) starting from the homogeneous phase seen at
    $\gamma=0$ in (b). They destroy in (f) the near-equilibrium phase
    separation shown in (e) that results from the attractive tail of
    the Lennard-Jones potential.}
  \label{fig:PhaseSep}
\end{figure}

In this Letter, we show that taking fluctuations into account leads to
a much richer scenario than reported so far and that nematic torques
can either induce or destroy phase separation in scalar systems, as
illustrated in Fig.~\ref{fig:PhaseSep} (all our simulations are
detailed in~\cite{supp}). Our central result is the discovery of the
underlying mechanism: the aligning nematic torques reduce, through
fluctuations, the mass of the polar field.  In turn, this enhances
polar order and particle persistence length, which plays an important
role in controlling the emergent properties.  We note that
fluctuations lower the critical temperature in aligning spin
systems~\cite{chaikin1995principles} so that this effect is the
opposite of what one would naively imagine. It is also the opposite of
what a phenomenological computation starting from a Landau theory
would predict. Instead, it relies on the precise correlations of the
fluctuations affecting polar and nematic fields.

To proceed, we construct the coupled stochastic field theory
describing the dynamics of the density, polar and nematic fields
emerging from Eq.~\eqref{eq:dynamics}. Using a weak-noise expansion,
we show that the correlated fluctuations of polar and nematic fields
conspire to lower the mass of the polar field. To test this
prediction, we consider the accumulation of self-propelled particles
against confining boundaries in the absence of pairwise
forces. Nematic torques then lead to an enhanced accumulation, which
is quantitatively accounted for by the renormalization of the
persistence length. Then, we consider repulsive forces between
particles and develop a new theory for the spinodal decomposition of
MIPS in the presence of aligning torques. We show that nematic
alignment enhances the active contribution to a generalized bulk
modulus, which becomes negative and induces MIPS for strong enough
alignment. \if{It also means that increasing the density of active
  particles in experiments should generically lead to an increase of
  their persistence due to nematically-aligning collisions, and not
  solely to a decrease of self-propulsion due to head-on
  collisions.}\fi Finally, we show that our results hold for more
general systems and that nematic torques enhance the persistence of
active particles in mixtures of active and passive particles as well
as in a nematic version of the Vicsek model. For mixtures, the
enhancement of the particle persistence length is strong enough to
suppress demixing. \if{Experimental active systems always involve a
  host of complex phenomena simultaneously and we hope that our study
  will help design active particles as well as account for their
  emerging properties.}\fi

\textit{Fluctuation-induced increase of the persistence length.}
Starting from the microscopic dynamics, Eq.~\eqref{eq:dynamics},
stochastic calculus allows deriving the time evolution of the
empirical measure $\hat \psi( \textbf{r}, \theta)= \sum_i \delta(
\textbf{r} - \textbf{r}_i) \delta( \theta - \theta_i)$, whose Fourier
modes $\hat f_k\equiv \int {\rm d} \theta \, \hat\psi(\bfr,\theta)
{\rm e}^{i k \theta}$ describe the fluctuating hydrodynamic fields of
the model. The particle density field indeed corresponds to $\hat
\rho(\bfr)=\hat f_0(\bfr)$ while $\hat f_1(\bfr)=\hat m_x(\bfr)+i \hat
m_y(\bfr)$ encodes the orientation field $\hat \bfm(\bfr)=\sum_i
\bfu(\theta_i)\delta(\bfr-\bfr_i)$.  The local nematic order is
quantified by $\hat f_2(\bfr)=2 {\hat q_{xx}}(\bfr)+i 2\hat
q_{xy}(\bfr)$~\cite{peshkov2014boltzmann}.  For simplicity, we set
$D_t=0$ but our results hold for finite $D_t$. Assuming that $\hat
\psi$ varies over scales larger than the interaction range $r_0$,
standard algebra leads to the following time evolution for $\hat
f_k$~\cite{supp}:
\begin{eqnarray}\label{eq:fourier}
\partial_t \hat f_k &+&\frac{v_0}2 (\nabla^* \hat f_{k+1}+\nabla \hat f_{k-1})+\mu\nabla\cdot \hat {\bf I}_k=\\
&&-k^2 D_r \hat f_k +\frac{k \gamma}{2 \hat f_0}(\hat f_2 \hat f_{k-2}
-\hat f_{-2}\hat f_{k+2})+\xi_k\,,\notag
\end{eqnarray}
where $\hat {\bf I}_k(\bfr)=\int {\rm d} \br' \, {\bf F}(\bfr-\bfr')
\hat f_k(\bfr) \hat f_0(\bfr')$. The $\xi_k$'s are centered Gaussian
white noise fields, induced by the individual noises $\zeta_i(t)$
entering Eq.~\eqref{eq:dynamicsangle}. They are given by
$\xi_k(\bfr,t)=i k \sqrt{2 D_r} \sum_i \zeta_i(t)
e^{ik\theta_i}\delta(\bfr-\bfr_i(t))$ and satisfy $\langle
\xi_k(\bfr,t)\xi_q(\bfr',t')\rangle_{\zeta_i}=\hat \Lambda_{kq}
\delta(t-t') \delta(\bfr-\bfr')$, with
\begin{equation}\label{eq:noisecorr}
        \hat \Lambda_{kq}=- 2 k q D_r \hat f_{q+k} \;.
\end{equation}
The bare mass of the orientation field $\hat f_1$ is thus equal to
$D_r$. The fluctuations, however, induce a non-vanishing correction to
this bare mass that we now compute.

To focus on the core mechanism, we consider the limit $v_0=\mu=0$ and
$r_0=\infty$ in Eq.~\eqref{eq:dynamics}, which amounts to studying $N$
fully-connected XY spins with nematic
alignment. Equation~\eqref{eq:fourier} directly generalizes to this
case, without the transport terms, with $\hat f_k(t)$ the $k$-th
Fourier mode of $\hat \psi(\theta)=\sum_i\delta(\theta-\theta_i)$, and
with $\hat f_0=N$. Therefore, the dynamics of the first two modes
read:
\begin{subequations}\label{eq:NXYf}
\begin{align} \dot{\hat f}_1
&= -D_r \hat f_1 + \frac{\gamma}{2N} \hat f_2 \hat f_{-1}
- \frac{\gamma}{2 N} \hat f_3 \hat f_{-2} + \xi_1 \label{eq:NXY-m}\,
,\\ \dot{\hat f}_2 &= -(4 D_r-\gamma) \hat f_2 - \frac{\gamma}{N} \hat
f_4 \hat f_{-2} + \xi_{2}\;. \end{align} \end{subequations}

When $N\to \infty$, the mean-field approximation to the dynamics
exactly predicts the relaxation of $f_k=\langle \hat f_k\rangle$. In
the high-temperature phase, the stable fixed points correspond to
$f_{k>0}^0 = 0$ and the masses of the polar and nematic fields are
$D_r$ and $4D_r-\gamma$, respectively.  To compute the first-order
correction to mean field, we evaluate $\frac{\gamma}{2N} \langle \hat
f_2 \hat f_{k-2}-\hat f_{-2} \hat f_{k+2}\rangle$ to leading order in
$N^{-1}$. Rewriting the Fourier modes as $\hat f_k=f_k^0+\delta f_k$
and using that $f_k^0=0$ in the high-temperature phase, one can expand the correlators to get
\begin{equation}
  \partial_t f_k  = - k^2 D_r f_k +\frac{k \gamma}{2N}\left( \mv{ \delta {f}_2  \delta {f}_{k-2}} -  \mv{ \delta{f}_{-2}  \delta{f}_{k+2} } \right) \: .
\end{equation}
Then, the linearized dynamics of the fluctuations in Fourier space
read $ \partial_t \delta f_k = - \left( k^2D_r - \gamma \delta_{|k|,
  2} \right) \delta f_k + \xi_k$. Using It\=o calculus, one then finds
the steady-state correlators
\begin{equation}\label{eq:NoiseCorrComplexGeneral}
\mv{\delta f_k \delta f_q} = \frac{ \Lambda_{kq}}{ (k^2+ q^2 )D_r - \gamma (\delta_{|k|,2} + \delta_{|q|,2} )} \: ,
\end{equation}
with $\Lambda_{kq}=\mv{\hat \Lambda_{kq}}$. This yields a renormalized dynamics for the modes given by
$\partial_t f_k = -\mathfrak{m}_k f_k +\mathcal{O}(f_k/N^2)$. For the
polar and nematic fields, we get the renormalized masses
\begin{subequations}\label{eq:renormalized_masses}
\begin{align}%
        \mathfrak{m}_1&=D_r - \frac{4\gamma D_r ( \gamma - D_r)}{N (5D_r - \gamma)(13D_r -\gamma)}+\mathcal{O}\left(\frac 1 {N^2}\right) \: ,\\
        \mathfrak{m}_2&=4D_r  - \gamma  + \frac{16D_r \gamma}{N ( 20D_r - \gamma)}+\mathcal{O}\left(\frac 1 {N^2}\right)\:.
\end{align}%
\end{subequations}%
As expected, fluctuations increase the mass of the nematic field,
shifting the ordering transition to temperatures lower than the
mean-field prediction $D_r=\gamma/4$.  Surprisingly, however,
Eq.~\eqref{eq:renormalized_masses} predicts that the mass of the polar
field is \textit{reduced} by fluctuations when $D_r<\gamma<5
D_r$~\footnote{Note that the computations leading to
Eq.~\eqref{eq:renormalized_masses} are valid only in the disordered
phase, where $\gamma \le 4D_r$, so that the spurious divergences
predicted in Eq.~\eqref{eq:renormalized_masses} are never observed.}.
All in all, fluctuations thus suppress the nematic order while they
favor the polar one.

While our results are perturbative, microscopic simulations reported
in Supplemental Material Fig.~S2 show these effects to hold
non-perturbatively. Note that our results rely on the exact expression
of the noise statistics, Eq.~\eqref{eq:noisecorr}, which we derived in
Eq.~\eqref{eq:fourier} from the microscopic dynamics,
Eq.~\eqref{eq:dynamics}. As shown in~\cite{supp}, complementing a
mean-field Landau theory by phenomenological uncorrelated noises would
(wrongly) lead to the opposite prediction of an increase of
$\mathfrak{m}_1$ due to fluctuations.

The interplay between fluctuations and nematic torques thus leads to a reduction of the polar field mass. This general result, which also holds in
equilibrium, implies that nematic torques enhance the persistence of
active particles. Together with the phase diagrams shown in
Fig.~\ref{fig:PhaseSep}, this qualitatively explains how phase
separation is either favored or suppressed by nematic torques. We now turn to
check the validity of our predictions as well as their scope. Before
considering the complex many-body dynamics of Eq.~\eqref{eq:dynamics}, we
consider a simpler problem in which persistence plays a key role.

\begin{figure}
  \centering
  \includegraphics{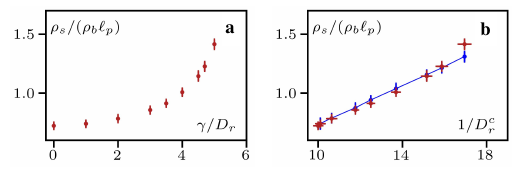}



  \caption{{\bf (a)} Ratio between boundary density and bulk density,
    $\rho_s/\rho_b$, normalized by the bare {persistence} $\ell_p$ as
    $\gamma$ is varied and the bare rotational diffusivity $D_r$ is
    kept constant. Measurements obtained from simulations of
    Eq.~\eqref{eq:dynamics} without pairwise forces. {\bf (b)} Red
    dots: same data as in panel a, plotted against
    $1/D_r^c(\gamma)$, where $D_r^c(\gamma)$ is extracted from the fit
    of $C_M(t)$. Blue dots connected by a line: Boundary accumulation
    of non-interacting particles whose bare rotational diffusivities
    $D_r$ are chosen equal to the measured values of $D_r^c(\gamma)$
    for the data of panel a, as a function of $1/D_r=1/D_r^c(\gamma)$.}
  \label{Fig:2}
\end{figure}

\textit{Boundary accumulation.} A typical trait of active particles is
their tendency to accumulate at confining
boundaries~\cite{elgeti2009self,tailleur2009sedimentation,fily2014dynamics,yang2014aggregation,bechinger2016active,deblais2018boundaries}
where they spend a typical time of order $\tau$ before escaping to the
bulk of the system. Dimensional analysis predicts that the ratio
between surface and bulk densities---which scale as inverse surface
and volume, respectively---should behave as $\rho_s/\rho_b\propto
v_0\tau=\ell_p$, where $\ell_p$ is the persistence length. In
Fig.~\ref{Fig:2}a, we show the results of simulations of
Eq.~\eqref{eq:dynamics} without interparticle forces, in the presence
of a confining potential. As $\gamma$ is increased up to $\gamma\simeq
5 D_r$, the fraction of particles at the walls increases by a factor
of 2 while the system remains disordered. Our analytical computations
suggest a simple qualitative explanation: aligning interactions lead
to a reduced effective rotational diffusivity $D_r^c$. In turn, this
yields an enhanced persistence length $\ell_p^c=v_0/D_r^c$ and thus an
increased boundary accumulation.

To test this hypothesis, we measured the auto-correlation function of
the total orientation,
$\hat\bfM(t)\equiv\int d\bfr \hat \bfm(\bfr,t)$, in the presence of
aligning torques:
$C_M(t)\equiv \langle \hat \bfM(t)\cdot \hat \bfM(0)\rangle$. For
$\gamma\leq 5 D_r $, the system remains disordered and $C_M(t)$ is
well fitted by an exponential decay $C_M(t)=\bfM_0^2 \exp(-D_r^c t)$,
from which we extract $D_r^c(\gamma)$. As predicted, $D_r^c(\gamma)$
is a decreasing function of $\gamma$.  We then compared the boundary
accumulation with that observed in simulations of
\textit{non-interacting} particles with rotational diffusivity
$D_r=D_r^c(\gamma)$. Remarkably, the excess densities are hardly
distinguishable below $\gamma=4.7 D_r$, i.e.\,when the interacting
system is far enough from the ordering transition
(Fig.~\ref{Fig:2}b). The renormalization of the polar-field mass,
which is a hydrodynamic effect, thus quantitatively accounts for the
particle accumulation at the boundaries, despite the microscopic
nature of this phenomenon.

\begin{figure}
  \centering
  \includegraphics{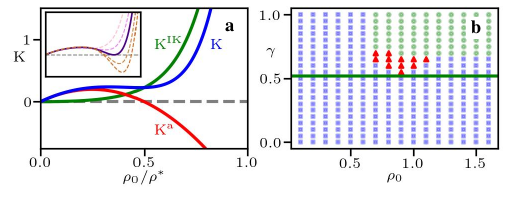}

  \caption{Onset of MIPS. {\bf (a)}
  Measurement of the active (red line) and passive (green line) components, ${\rm K}^{\rm a}$ and ${\rm K}^{\rm IK}$ respectively, of the generalized bulk modulus ${\rm K}$ (blue line) as the density $\rho_0$ is varied, in the absence of nematic torques. The
  density is normalized by $\rho^*$, the density at which the
  effective self-propulsion speed $v(\rho)\equiv \langle \dot {\bf
    r}_i\cdot \bfu(\theta_i)\rangle$ vanishes.
    From the measurement of $\mass_1(\rho_0, \gamma)$, Eq.~\eqref{eq:predictK} predicts the evolution of ${\rm K}^{\rm th}(\rho_0, \gamma)$ when $\gamma$ is increased. Five representative curves are shown in the inset, with $\gamma$ ranging from 0 to 0.6. The solid line corresponds to $\gamma_c=0.52$.
    {\bf
      (b):} Phase diagram corresponding to simulations of
    Eq.~\eqref{eq:dynamics} as $\rho_0$ and $\gamma$ are
    varied. Blue squares correspond to homogenous disorder systems. Red
    triangles correspond to the MIPS region. Green circles correspond
    to the emergence of nematic order. The solid green line corresponds to the theoretical
    prediction $\gamma_c=0.52$.}\label{fig:pressure}
\end{figure}

\textit{Nematic torques and MIPS.} Let us now show that the
renormalization of the polar-field mass also quantitatively accounts
for the emergence of MIPS at finite $\gamma$. To do so, we first
derive the relaxation dynamics of the density field. For $k=0$,
Eq.~\eqref{eq:fourier} reads $\dot \rho = - \nabla\cdot \bfJ$, where
$\bfJ(\bfr)=v_0 \bfm (\bfr)+\mu\bfI_0(\bfr)$, $\bfm=\langle
\hat\bfm\rangle$, and $\bfI_0(\bfr)=\langle \int {\rm d} \bfr' \,
\hat\rho(\bfr) \bfF (\bfr-\bfr')\hat\rho(\bfr')\rangle$. The dynamics
of $\bfm$ then stems from that of $f_1$ as
\begin{equation}
  \partial_t \bfm = - \nabla \cdot \Big [ v_0 \Big(\bfQ +  \frac{\rho\mathbb{I}}{2}\Big ) + \mu\bfI_{1} \Big] - D_r \bfm +\Big\langle \frac{\gamma}{\hat \rho} \hat \bfQ \cdot \hat \bfm - \frac{2\gamma}{\hat \rho} \hat \bfchi\cdot\hat \bfQ\Big\rangle \label{eq:polar}
\end{equation}
where $\mathbb{I}_{\alpha\beta}=\delta_{\alpha\beta}$,
$\bfI_{1,\alpha\beta}(\bfr)=\langle \int {\rm d} \bfr' \, \hat
m_\beta(\bfr) F_\alpha (\bfr-\bfr')\hat \rho(\bfr')\rangle$,
$\hat\bfQ_{\alpha\beta}(\bfr)=\sum_i (\bfu_{i,\alpha}
\bfu_{i,\beta}-\frac{\delta_{\alpha\beta}}2)\delta(\bfr-\bfr_i)$ is
the nematic order field, $\bfQ=\langle \hat\bfQ\rangle$, and
$\hat{\bfchi}_{\alpha \beta \gamma}$ is a third-order
tensor~\cite{supp}.  In general, closing Eq.~\eqref{eq:polar} for the
field $\bfm$ is a difficult task. In light of our results, we predict
that, in the high-temperature disordered phase, the non-linear
aligning terms can simply be accounted for by a renormalization of the
bare mass $D_r$ of the polar field. The non-conserving terms in
Eq.~\eqref{eq:polar} thus reduce to $-\mass_1(\rho,\gamma) \bfm$. A fast
variable treatment on $\bfm$ then allows us to rewrite the dynamics of
$\rho$ as
\begin{equation}\label{eq:dynrho}
\dot {\rho} = - \nabla\cdot \Big[\mu\frac{ D_r}{\mass_1} \nabla \cdot \bsigma^{\rm a}+ \mu \nabla \cdot \bsigma^{\rm IK}\Big],
\end{equation}
where we have introduced $\bsigma^{\rm a}= - v_0^2 (\bfQ +
\frac{\rho\mathbb{I}}{2})/(\mu D_r) - v_0 \bfI_{1}/D_r$ and followed Irving
and Kirkwood to rewrite the contribution of pairwise forces as
$\bfI_0= \nabla\cdot \bsigma^{\rm
  IK}$~\cite{irving_statistical_1950,kruger2018stresses}. To assess
the stability of an isotropic, homogeneous phase at density $\rho_0$,
we compute the linearized dynamics in Fourier space of a fluctuation
along, say, the $\hat x$ axis, which reads
\begin{equation}\label{eq:linstab}
  \partial_t \delta\rho_q = \mu q^2
  \Big[\frac{D_r}{\mass_1}{\sigma_{xx}^{\rm
        a}}'(\rho_0)+{\sigma_{xx}^{\rm IK}}'(\rho_0)\Big]
  \delta\rho_q\;,
\end{equation}
where the prime denotes derivative with respect to $\rho_0$.

When $\gamma=0$, $\bsigma^{\rm a}$ is the contribution of the active
forces to the stress
tensor~\cite{takatori2014swim,yang2014aggregation,solon2015pressure},
$\mass_1=D_r$, and Eq.~\eqref{eq:dynrho} reduces to $\dot \rho = -
\nabla \cdot (\mu \nabla \cdot \bsigma)$~\cite{solon2015pressure}. The
mechanical pressure exerted by active particles then satisfies an
equation of state given by $P=-\text{Tr}\bsigma/2$ and an isotropic
homogeneous profile at density $\rho_0$ is linearly unstable whenever
$P'(\rho_0)=-{\sigma_{xx}^{\rm a}}'(\rho_0)-{\sigma_{xx}^{\rm
    IK}}'(\rho_0)<0$. This is the standard mechanical route to account
for the MIPS induced by repulsive pairwise
forces~\cite{solon2015pressure,solon2018generalized,solon2018njp,speck2021coexistence,omar2023mechanical}. The
system is thus unstable when its bulk modulus is negative: ${\rm
  K}=\rho_0 P'(\rho_0)={\rm K}^{\rm a}(\rho_0)+ {\rm K}^{\rm
  IK}(\rho_0)<0$, where $ {\rm K}$ has been split into active and
passive components ${\rm K}^{\rm a}(\rho_0)\equiv
-\rho_0{\sigma_{xx}^{\rm a}}'(\rho_0)$ and ${\rm K}^{\rm IK}(\rho_0)
\equiv -\rho_0 {\sigma_{xx}^{\rm IK}}'(\rho_0)$.

In the presence of aligning torques, despite the lack of equation of
state~\cite{solon2015nphys}, $ {\rm K}^{\rm a}$ and $ {\rm K}^{\rm
  IK}$ still control the stability of homogeneous profiles through
Eq.~\eqref{eq:linstab}. This suggests defining a `generalized bulk
modulus'---without connection to mechanics---as $ {\rm K}\!\equiv\!
\frac{D_r}{\mass_1} {\rm K}^{\rm a} + {\rm K}^{\rm IK}$. Like in
equilibrium, negative values of $K$ then lead to a spinodal
decomposition. In the disordered phase, we expect that $\gamma$ barely
alters the values of $ {\rm K}^{\rm a}$ and $ {\rm K}^{\rm IK}$ (see Supplemental Material Fig.~S3a-b). Their measurements at $\gamma=0$, show that $ {\rm
  K}^{\rm a}$ favors instability whereas $ {\rm K}^{\rm IK}$ stabilizes
homogeneous phases (see Fig~\ref{fig:pressure}a). Their sum is
positive and the system is stable. As $\gamma$ increases, we estimate
the generalized bulk modulus as
\begin{equation}\label{eq:predictK}
   {\rm K}^{\rm th}(\rho_0,\gamma)\equiv  {\rm K}^{\rm
    IK}(\rho_0)+\frac{D_r}{\mass_1(\rho_0,\gamma)}  {\rm K}^{\rm
    a}(\rho_0)\;.
\end{equation}
The renormalization of $\mass_1$ thus enhances the contribution of
$ {\rm K}^{\rm a}$ by a factor of $\frac{D_r}{\mass_1}$. The inset of
Fig~\ref{fig:pressure}a shows $ {\rm K}^{\rm th}(\rho_0,\gamma)$ for several
values of $\gamma$.  For $\gamma>\gamma_c=0.52$, the active bulk
modulus dominates and we predict the occurence of MIPS.
This is successfully compared with simulations of
Eq.~\eqref{eq:dynamics} in the ($\gamma,\rho_0$) plane in
Fig.~\ref{fig:pressure}b. All in all, the renormalization of $\mass_1$
due to the nematic torques thus induces MIPS by increasing the active
contribution to the bulk modulus.

\begin{figure}
  \centering
  \includegraphics{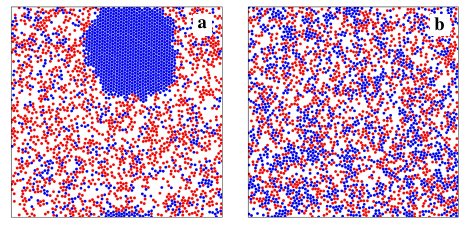}
    \caption{Impact of nematic torques on mixtures of active particles
      (red circles) and passive particles (blue circles) evolving
      under Eqs.~\eqref{eq:dynamicsposact}
      and~\eqref{eq:dynamicspospass}, respectively. {\bf (a)} In the
      absence of nematic torques, passive particles undergo phase
      separation whenever the persistence length of the active
      particles is small enough. {\bf (b)} Nematic alignment between
      the active particles ($\gamma=4.5D_r$) suppresses the phase
      separation of the passive component.}
     \label{fig:mixture}
  \end{figure}

\textit{Mixture of active and passive particles.} To show that our
results apply more broadly, we consider mixtures of active and passive
particles interacting via purely repulsive forces, which have
attracted a lot of attention
recently~\cite{stenhammar2015activity,takatori2015theory,weber2016binary,wysocki2016propagating,wittkowski2017nonequilibrium,dolai2018phase,alaimo2018microscopic}. Active
particles are characterized by positions $\bfr^{\rm a}_i$ and orientations
$\bfu(\theta_i)$ while the positions of passive particles are denoted
by $\bfr^{\rm p}_i$. The spatial dynamics read
\begin{subequations}\label{eq:dynamicsmixtures}
\begin{align}
  \dot \bfr_i^{\rm a} &= v_0 \bfu_i -\mu \!\!\!\!\!\!\!\!\sum_{|\bfr_i^{\rm a}-\bfr_j^{\rm a}|<r_0} \!\!\!\!\!\!\!\!\nabla U(\bfr_i^{\rm a}-\bfr_j^{\rm a})-\mu \!\!\!\!\!\!\!\!\sum_{|\bfr_i^{\rm a}-\bfr_j^{\rm p}|<r_0} \!\!\!\!\!\!\!\! \nabla U(\bfr_i^{\rm a}-\bfr_j^{\rm p}) \label{eq:dynamicsposact}\\
  \dot \bfr_i^{\rm p} &= -\mu \!\!\!\!\!\!\!\!\sum_{|\bfr_i^{\rm p}-\bfr_j^{\rm p}|<r_0} \!\!\!\!\!\!\!\! \nabla U(\bfr_i^{\rm p}-\bfr_j^{\rm p})-\mu \!\!\!\!\!\!\!\!\sum_{|\bfr_i^{\rm p}-\bfr_j^{\rm a}|<r_0} \!\!\!\!\!\!\!\! \nabla U(\bfr_i^{\rm p}-\bfr_j^{\rm a}) \label{eq:dynamicspospass}\;,
\end{align}
\end{subequations}
and the orientations of the active particles evolve according to
Eq.~\eqref{eq:dynamicsangle}. In such mixtures, the passive component
undergoes phase separation when the persistence length of the active
particles is small enough~\cite{weber2016binary}, as shown in the
simulation reported in Fig.~\ref{fig:mixture}a for $\gamma=0$. Our
results suggest that increasing $\gamma$ should enhance the
persistence of the active particles, thereby vaporising the dense
passive phase. This is indeed what is reported in
Fig.~\ref{fig:mixture}{b} where $\gamma$ has been increased up to
$\gamma/D_r\simeq 4.5$. This shows that the impact of nematic torques
in disordered scalar matter can be generically accounted for by an
increase of the persistence length.

\textit{Conclusion.} We have shown how fluctuations enhance polar
order in collections of nematically aligning particles. In active
systems, this leads to an increase of the persistence length, hence
impacting emergent properties.  As suggested by our hydrodynamic
treatment, we expect this effect to be generic and to extend beyond
systems described by Eq.~\eqref{eq:dynamics}. Supplemental Material Sec. V for
instance show how nematic torques also enhance persistence in
discrete-time Vicsek-like models~\cite{supp}. Similar results should
also apply to passive systems where, to the best of our knowledge, it
has not been reported before. In experiments, increasing the density
of active particles should generically lead to an increase of their
persistence due to nematically aligning collisions, which should
oppose the well-documented decrease of self-propulsion due to head-on
collisions~\cite{fily2012athermal}. Note that our work uses the same
range for pairwise forces and aligning torques, hence being as close
as possible to collision-based models, where repulsive forces and
aligning dynamics go hand in hand. For swarming bacteria like
\textit{B. subtilis}, we suspect that a difference in shapes between
bacteria at the fore-front and in the core of the colony could
regulate the bacterial persistence and play an important role in the
spreading of the swarm.

\textit{Acknowledgments:} We thank Josep-Maria Armengol-Collado, François Graner,
Yariv Kafri, Sunghan Ro, Alex Solon, and Fred van Wijland for
stimulating discussions. J.T., C.D., and G.S. acknowledge support from the
ANR grant THEMA. C.D. acknowledges the support of a postdoctoral
fellowship from the LabEx ``Who Am I?''  (ANR-11-LABX-0071) and the
Université Paris Cité IdEx (ANR-18-IDEX-0001) funded by the French
Government through its ``Investments for the Future'' program.

\nocite{dean1996langevin,doi1981molecular,martin2021statistical,patelli2019understanding,kursten2021quantitative,chou2015active,martin2021fluctuation}

\bibliographystyle{apsrev4-2}
\bibliography{biblio}

\end{document}